\RequirePackage[T1]{fontenc}
\documentclass[journal]{IEEEtran}
\usepackage[letterpaper,margin=1in,bottom=1.05in,includehead]{geometry}

\usepackage{epsfig}
\usepackage{graphicx}
\usepackage{amsmath}
\usepackage{amssymb}
\usepackage{booktabs}
\usepackage{multirow}
\usepackage{tikz}
\usepackage{xcolor}
\usepackage{cite}
\usepackage{placeins}

\definecolor{ieeeblue}{rgb}{0.10,0.32,0.55}
\usepackage[breaklinks,colorlinks,allcolors=ieeeblue]{hyperref}
\hypersetup{
  pdftitle={Neural Grey: Practical Vignetting Calibration from Casual Camera Rolls},
  pdfauthor={Yueze Wang and Spencer LaVere Smith},
  pdfsubject={Vignetting calibration},
  pdfkeywords={Neural Grey, vignetting calibration, lens shading correction}
}


\begin{document}
\raggedbottom

\title{Neural Grey: Practical Vignetting Calibration from Casual Camera Rolls}

\author{Yueze Wang and Spencer LaVere Smith\thanks{The authors are with the
Department of Electrical and Computer Engineering, University of California,
Santa Barbara, Santa Barbara, CA, USA (e-mail: yueze@ucsb.edu;
sls@ucsb.edu). Corresponding author: Spencer LaVere Smith.}}
\maketitle

\begin{abstract}
Neural Grey estimates a reusable camera--lens shading map from four images of
a static, light-colored surface taken at a range of arbitrary roll angles.
Rolling the camera moves wall
illumination across the sensor while lens shading stays fixed. We fit these
two fields jointly, together with small pose and exposure errors, using a
physical shading envelope and compact bounded neural residuals. Rotation alone,
however, cannot distinguish radial lens attenuation from radial wall
illumination. A focal-conditioned radial selector addresses this ambiguity:
it uses focal length and sensor size to choose the broad radial profile while
retaining the fitted spatial remainder. Component and sensitivity studies
separate the roles of the field representation, radial selection, and numerical
settings. On 100 synthetic cases, mean peak signal-to-noise ratio (PSNR) for
the estimated shading maps is 25.43~dB for Neural Grey and 23.72~dB for
four-shot DeVigNet. Neural Grey recovers one ray-traced lens profile exactly
and averages 24.49~dB on the other 11. The closest competing method (Kim and
Pollefeys) averages 21.95~dB across all 12. On 21 real takes, Neural Grey
averages 25.38~dB, exceeds the strongest competing mean by 5.62~dB, and wins
five of seven optical configurations.

\end{abstract}


\section{Introduction}
Vignetting makes image brightness depend on position as well as scene
radiance. This complicates photometric measurement, scientific imaging, and
consistent correction when lenses are changed
\cite{debevec1997hdr,smith2015cidre}.
The effective shading response depends on the camera, lens, aperture, and focus
state \cite{hecht2016optics,cooke2021itech}. We seek a reusable calibration map for one such configuration.

Controlled flat-field captures remain the most accurate option when suitable
lighting and equipment are available
\cite{zheng2009vignetting,yu2004vignetting,kask2016flatfield}.
In the field, however, repeating this calibration can be inconvenient.
Analytic profiles and database lookup tables are efficient, but can miss
unit-specific decentering, sensor response, or an unlisted lens
\cite{hecht2016optics,cooke2021itech}. Large restoration networks offer more
flexibility, but four calibration captures cannot provide the training data
normally required by a population model.

Our starting point is a simple capture experiment: roll the camera while
viewing the same light-colored wall. Wall illumination moves across the sensor;
camera--lens shading stays in place. Fitting the captures together can therefore
separate much of their spatial structure. There is one essential limitation:
a radial darkening looks unchanged after a roll, so the images alone cannot tell
whether it belongs to the lens or the wall. We call this a radial gauge ambiguity.

Neural Grey follows this distinction throughout its design. A cosine-power
base and small bounded neural fields fit the shading and illumination without
freely reproducing wall texture. A subsequent radial selector uses focal length
and sensor size to choose the broad attenuation profile, retaining the spatial
remainder from the fitted captures. The output is a reusable shading map for
one camera--lens--aperture--focus configuration. Our experiments separately test
the field representation, the metadata-based radial choice, and the capture
protocol.

Our contributions are:
\begin{itemize}
\item A joint radiometric model that separates sensor-fixed shading from
scene-fixed illumination and states the spatial ambiguity left by camera rolls.
\item A bounded physical--neural representation and a focal-conditioned
radial selector that supplies prior information absent from the roll captures.
\item Evaluation on synthetic profiles, ray-traced optics, real captures, and
measured-profile stress cases, with component, simplification, sensitivity,
and capture-count tests that distinguish the roles of the design choices.
\end{itemize}

\begin{figure*}[!t]
\centering
\includegraphics[width=\textwidth]{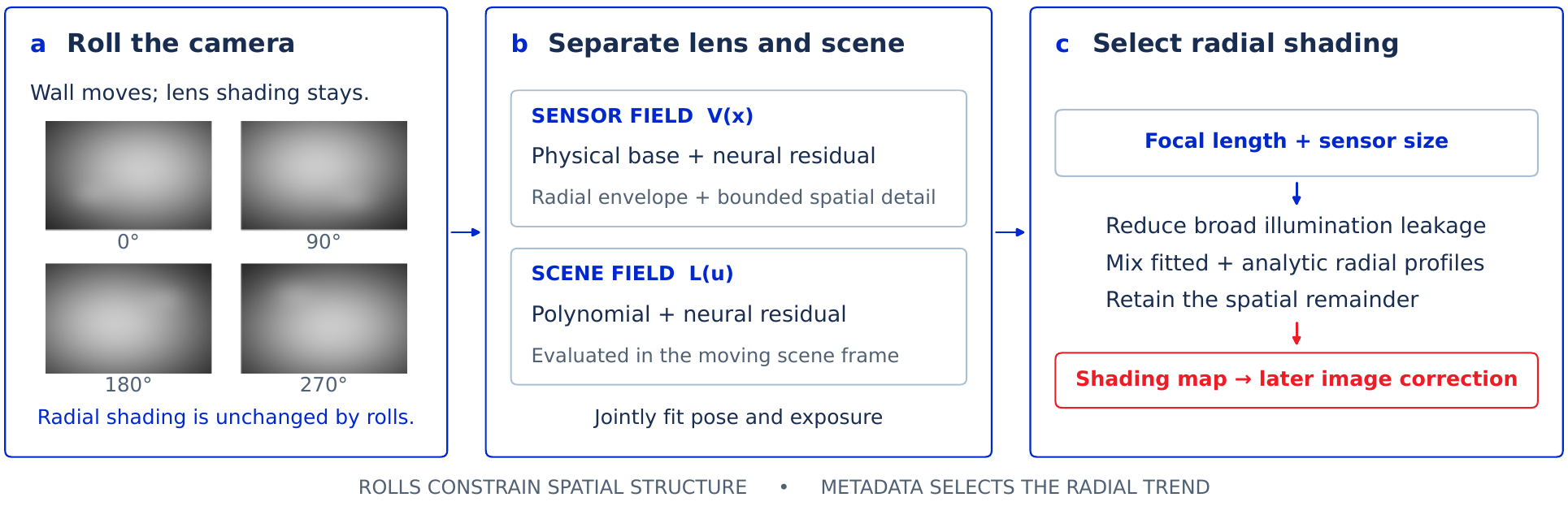}
\caption{Neural Grey follows what the captures can and cannot reveal.
(a) Camera rolls move wall illumination relative to sensor-fixed shading.
(b) A physical base and bounded neural residual jointly fit the shading,
scene illumination, pose, and exposure. (c) Focal metadata selects the ambiguous
radial profile after low-frequency stabilization; the remaining spatial detail
is retained. The resulting map corrects later images at the same optical
configuration. Field maps illustrate the mechanism, not measured results.}
\label{fig:architecture}
\end{figure*}

\section{Related Work}

\subsection{Physical and Geometric Correction}
Classical correction fits cosine-power, elliptical, or even-polynomial radial
profiles \cite{hecht2016optics,zheng2009vignetting}, sometimes with a learned
optical-center shift \cite{kang2000vignetting,kim2008vignetting}. These models
are efficient and interpretable, but near-radial assumptions cannot capture all
unit-specific decentering, sensor-stack response, or asymmetric structure
\cite{goldman2010vignette}.
Retrospective microscopy correction instead estimates shared shading from image
content, using entropy minimization \cite{likar2000entropy} or multi-image
models \cite{smith2015cidre,peng2017basic}.
Radiometric calibration also estimates camera response from exposure changes
\cite{mitsunaga1999radiometric,grossberg2003knowable}, image collections
\cite{mo2017radiometric}, or faces \cite{li2017faces}.

\subsection{Purely Data-Driven Models}
Learned correction includes UNet-style models
\cite{ronneberger2015unet,chen2018cnn}, lookup-table (LUT) formulations such as VCLUT
\cite{vclut2023}, frequency-aware DeVigNet \cite{devignet2024}, and discrete
Gaussian-filter estimation \cite{abdelhamed2022filters}. Such population models
typically require representative training data and predict a corrected image.
Our per-configuration setting instead estimates a reusable shading field from
four captures; Sec.~V compares both pretrained and per-instance learned baselines.
Related illumination work includes classical Retinex models
\cite{land1971retinex,jobson1997multiscale}, learned decomposition
\cite{wei2018retinex,zhang2019kind}, reference-free curve and illumination
estimation \cite{guo2020zerodce,liu2021ruas,ma2022sci}, and noise-aware
enhancement \cite{xu2022snr}. These methods address image enhancement rather than reusable
lens calibration, but share the need to distinguish illumination from image
content.

\subsection{Physics-Anchored Neural Fields and Inverse Imaging}
Physics-informed learning places governing structure inside an inverse problem
\cite{raissi2019pinn,karniadakis2021piml}, while implicit neural fields provide
compact continuous parameterizations
\cite{mildenhall2020nerf,sitzmann2020siren}. Related representations encode
3D shape \cite{park2019deepsdf,mescheder2019occupancy}, continuous images
\cite{chen2021liif}, and illumination-related image transformations
\cite{yang2023nerco}. Image-specific network fitting can
also act as a prior without external training
\cite{ulyanov2018dip,gandelsman2019doubledip}. Neural Grey combines these tools
in a roll-capture factorization and explicitly treats its radial ambiguity.

\section{Proposed Method}
\label{sec:method}

\subsection{Problem Formulation}
Figure~\ref{fig:architecture} summarizes the method. We first fit all captures
with separate sensor and scene fields, then use focal metadata to select their
unresolved radial component. The distinction matters: a good reconstruction of
the four inputs is necessary, but does not by itself guarantee correct shading.

Let $x$ denote a two-dimensional sensor coordinate and $u$ denote coordinates on the
static calibration surface. The unknown $V(x)>0$ is fixed to the sensor, while
the surface radiance and illumination field $L(u)>0$ is fixed to the scene.
Following standard radiometric image formation models
\cite{debevec1997hdr,grossberg2004response,goldman2010vignette}, capture $k$ obeys
\begin{equation}
    I_k(x) = g_k\,V(x)\,L(W_kx) + b_k+n_k(x),
    \label{eq:image_formation}
\end{equation}
where $k\in\{1,\ldots,4\}$ indexes sparse, casual, approximate quarter-turn
rolls. The similarity warp
$W_kx=s_kR_{\alpha_k}x+t_k$, gain $g_k$, black-level offset $b_k$, and
zero-mean capture noise $n_k$ are nuisance variables; all except $n_k$ are
jointly estimated. Out-of-bounds
coordinates are removed by an overlap mask. We use four approximately
quarter-turn captures of a static, light-colored, approximately Lambertian
surface. Intensities are linearized; scene-dependent camera
processing can otherwise violate this model \cite{nam2017pipeline}.

In practical terms, a mark or lighting gradient on the wall moves to different
sensor pixels after a roll, whereas lens shading stays at the same sensor
pixels. Equation~\eqref{eq:image_formation} uses this difference to separate the
two fields while accounting for small hand-held and exposure errors.

\noindent\textbf{Proposition 1 (ambiguity under camera rotations).}
Assume $W_k=R_k$ are exact rotations about a common optical center. For every
positive field $Q$ invariant to all measured rotations,
$Q(R_kx)=Q(x)$, the transformation
\begin{equation}
 V'(x)=V(x)Q(x),\qquad L'(u)=L(u)/Q(u)
 \label{eq:radial_gauge}
\end{equation}
leaves all noiseless observations in Eq.~\eqref{eq:image_formation} unchanged.

{\raggedright\noindent\emph{Proof.} By the stated invariance,
$V'(x)L'(R_kx)=V(x)Q(x)L(R_kx)/Q(R_kx)=V(x)L(R_kx)$ for every $k$.\par}

Thus, any pattern that looks unchanged under all measured rolls can be moved
between $V$ and $L$ without changing the images. For ideal quarter turns this
includes every positive $C_4$-invariant field, and therefore every radial field
$Q(x)=q(\lVert x\rVert)$. Mean-centering $\log L$ fixes only the scalar scale;
it does not fix this spatial ambiguity. Bounded fields stabilize the observable
part, and independent focal metadata selects the dominant radial part.

\subsection{Bounded Shading and Illumination Fields}
We model sensor-fixed shading
as a smooth radial base with a bounded spatial correction:
\begin{equation}
\begin{aligned}
V_{phys}(x)&=I_0\cos^p\!\left(\arctan(r_{ell}/f_{norm})\right),\\
V_{pred}(x)&=V_{phys}(x)\,[1+\Delta V(x)].
\end{aligned}
\label{eq:bounded_field}
\end{equation}
The elliptical radius $r_{ell}$ uses a learned center, axis scales, and rotation
to accommodate decentering and unequal horizontal/vertical falloff. The exponent
$p\in[2,6]$ and positive scale $f_{norm}$ control the base profile; $f_{norm}$ is
an optimization variable, distinct from the recorded focal length used later.

A compact coordinate multilayer perceptron (MLP) with smooth Tanh activations
represents departures from the physical base. Its output is bounded and
complemented by low-order angular structure:
\begin{equation}
\Delta V(x)=\alpha_V\tanh(f_\eta(z))+e(r_{ell})\mathcal H_\omega(\phi).
\label{eq:spatial_residual}
\end{equation}
Here $z$ combines position, elliptical radius, and angular features, while
$\mathcal H_\omega$ contains bounded second- and fourth-order harmonics.
The radial envelope $e$ suppresses angular terms near the center, where
direction is ill-defined. Both branches are zero-initialized, so fitting
starts from the physical field. The bounded correction accommodates spatial
departures while limiting the freedom to reproduce wall texture. The
Supplementary Material specifies the features, the residual cap
$\alpha_V=0.18$, and the fixed network dimensions.

The scene illumination field
uses a second-order polynomial for tilt and broad curvature, plus a compact
Tanh MLP for smooth structure left by the polynomial:
\begin{equation}
\begin{aligned}
\ell(u)&=q_2(u)+0.25\tanh(h_\psi(u)),\\
\log L(u)&=\ell(u)-\operatorname{mean}_{u}\ell(u).
\end{aligned}
\end{equation}
Its neural term permits a multiplicative variation of approximately
$[0.78,1.28]$ before mean-centering. Unlike $V(x)$, this field is evaluated at
the warped scene location $u=W_kx$ in each capture. The separate coordinate
frames supply the cross-roll constraint; the polynomial, bounded residuals,
and smooth Tanh activations limit what either field can absorb.
We omit high-frequency Fourier encoding to favor smooth fields
\cite{rahaman2019spectral,tancik2020fourier}; explicit bandwidth control is
another approach to this representation issue \cite{lindell2022bacon}.

The resulting model contains approximately 24.8K parameters. Its capacity is
kept fixed across optical configurations; Sec.~\ref{sec:ablation} examines
whether nearby network sizes or simpler field representations improve recovery.

\subsection{Joint Fitting and Scale Normalization}

The loss has three roles: the data
term reconstructs all four images, the field terms limit how much structure
$V$ and $L$ can absorb, and the nuisance terms keep gains, warps, focal scale,
and center near their acquisition priors. We jointly optimize the
shading parameters $\{p,f_{norm},x_c,y_c,a_x,a_y,\varphi\}$, per-capture nuisance
variables $\{g_k,b_k,W_k\}$, and the polynomial-plus-MLP field $L$:
\begin{equation}
\begin{aligned}
    \mathcal{L} =&\ \mathcal{L}_{data}
    + \lambda_{phys} \mathcal{L}_{phys}\\
    &+ \lambda_{light} \mathcal{L}_{light}
    + \lambda_{anchor} \mathcal{L}_{anchor}\\
    &+ \lambda_{res} \mathcal{L}_{res}
    + \lambda_{gb}\mathcal L_{gb}\\
    &+ \lambda_{warp}\mathcal L_{warp}
    +\lambda_f\mathcal L_f+\lambda_c\mathcal L_c.
\end{aligned}
\end{equation}
The data term enforces consistency across all four captures with
$w_e(x,y)=1+1.25r^2$, normalized to unit mean:
\begin{equation}
    \mathcal{L}_{data} = \frac{1}{4} \sum_{k} w_e \left\| I_k - g_k \cdot V_{pred} \cdot L(W_kx) - b_k \right\|^2_2.
\end{equation}
Before normalization this raises the corner contribution from $1$ to $2.25$,
countering the concentration of image area at moderate radii and emphasizing
where attenuation provides the strongest signal.
Here $\mathcal L_{phys}=\|V_{pred}-V_{phys}\|^2$ retains the physical base,
$\mathcal L_{light}=\|\log L\|^2$, and
$\mathcal L_{res}=\|\Delta V\|^2$. Constructed by robust fusion of the
exposure-normalized, registered inputs, $V_{fused}$ supplies an internal
self-supervisory anchor independent of any external flat-field reference.
We use
$\mathcal L_{anchor}=\mathrm{SmoothL1}(V_{pred},V_{fused};\beta=0.03)$
\cite{huber1964robust}; the 3\%-transmittance transition is quadratic for small
fusion errors and linear for larger wall or registration remnants. Furthermore,
$\mathcal L_{gb}=\|g-1\|^2+\|b\|^2$,
$\mathcal L_{warp}$ is the squared translation, rotation, and log-scale
departure, $\mathcal L_f$ is the squared relative departure from the
metadata-derived normalized focal length, and $\mathcal L_c$ penalizes motion
of the optical center away from its robust initialization.

We average field losses over sampled coordinates and nuisance penalties over
captures. The same eight regularization weights are used for all datasets and
lenses. They balance reconstruction accuracy, plausible fields, and small
acquisition errors. The Supplementary Material lists their values and tests each at half
and twice its default value to assess local sensitivity.

Mean-centering $\log L$ fixes
the global scale between shading and illumination. Bounded per-capture gains
account for exposure differences. Exact bounds and initialization are given
in the Supplementary Material.

All parameters are jointly optimized using Adam \cite{kingma2015adam} with
cosine annealing \cite{loshchilov2017sgdr}. The neural residual uses the fastest
learning rate; global optical, illumination, and exposure variables change more
slowly, allowing local field corrections to adapt within a stable optical
and geometric fit. The Supplementary Material gives the fixed optimization
schedule, coordinate sampling, learning rates, and parameter bounds.

\subsection{Selecting the Remaining Radial Profile}
The joint fit returns $\widehat V$, but the capture model cannot determine its
broad radial trend uniquely. We therefore make a separate, explicit choice
using focal length and sensor size. This independent information about the
optical configuration complements the spatial constraints supplied by camera rolls.

Before selecting a radial profile,
we reduce large one-sided, low-frequency structure that can leak from the wall
into the shading estimate. A reflection average $\mathcal P_{D_2}$ averages the
smoothed map with its horizontal, vertical, and two-axis reflections. A soft
gate applies little correction when the asymmetry is small and stronger
correction when it is large. Applied to the low-frequency component, this
$D_2$ prior suppresses one-sided illumination leakage while retaining the
fine-scale remainder. Its preference for reflection symmetry can also attenuate
genuine broad optical asymmetry.

We decompose the stabilized
log map into a radial-bin median $p_b(r)$ and a spatial remainder $a(x)$.
Focal length and sensor diagonal define two analytic cosine-power profiles:
$p_w$ represents stronger wide-angle falloff, while $p_s$ provides a second
profile for standard and longer focal lengths. Continuous weights depend
on focal length and disagreement with the fitted profile:
\begin{equation}
\begin{aligned}
p_{out}(r)&=(1-\beta_w-\beta_s)p_b(r)
             +\beta_wp_w(r)+\beta_sp_s(r),\\
V_{out}(x)&=\mathcal N\!\left\{\exp[a(x)+p_{out}(r(x))]\right\},
\end{aligned}
\label{eq:radial_head}
\end{equation}
where $\mathcal N$ sets the peak to one. This replaces only the radial profile
of the stabilized map. It neither re-fits the neural fields nor queries a lens
database. Without focal metadata, we use the stabilized estimate.

The selector operates in normalized sensor coordinates, using radial medians
to separate broad attenuation from the spatial remainder. Gaussian smoothing
has a standard deviation of 3.9\% of the image long edge, and the radial
profile is discretized into 48 bins. Soft focal transitions are anchored at
35 and 42~mm on a 35-mm-equivalent scale. These optical priors address the
ambiguity identified by Proposition~1; their numerical values are empirical
development choices held fixed across evaluation. The Supplementary Material
gives the complete rule and tested sensitivity ranges.

\section{Model Analysis and Ablation Studies}
\label{sec:ablation}

The analysis separates representation capacity, radial ambiguity, and capture
robustness. A 15-scenario representation study, comprising nine synthetic and
six separate measured-profile scenarios, tests removal of the physical and
neural components (Table~\ref{tab:ablation}). A 12-case diagnostic benchmark
then evaluates radial selection, capture count, and local sensitivity
(Tables~\ref{tab:selector_gauge_ablation}--\ref{tab:tolerance}). It comprises
nine Lensfun-based and three analytic optical profiles with simulated
illumination and noise. Within each study, we compare estimated shading maps
with reference maps using peak signal-to-noise ratio (PSNR). We also report
the structural similarity index (SSIM) for the 12-case benchmark. SSIM
compares the brightness, contrast, and structure of the two maps, with
values near 1 indicating strong agreement.

\begin{figure*}[!t]
\centering
\includegraphics[width=\textwidth]{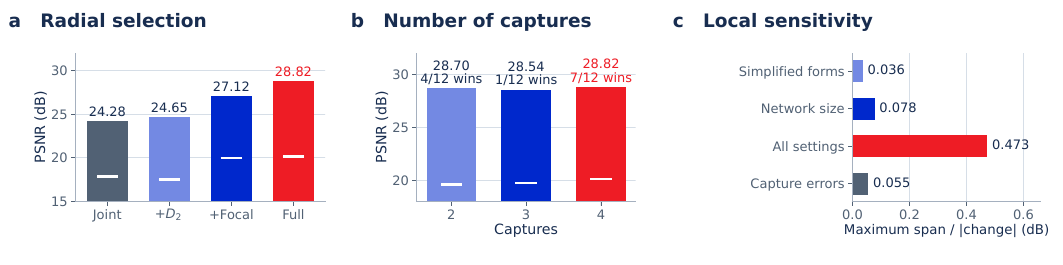}
\caption{Design diagnostics on the 12-case diagnostic benchmark. (a) Selector variants
on fixed four-capture fits. (b) Refits using the first two, three, or four captures. Bars show
mean PSNR; white marks show the worst case. (c) Maximum mean-PSNR spans across
simplified forms, network sizes, loss weights, and representation settings; capture errors
show the largest absolute endpoint change. Panels (a,b) use the full fitting schedule; panel (c) uses shorter fits
for local sensitivity. Capture-error sweeps average three scene realizations.
The Supplementary Material gives the numerical settings.}
\label{fig:design_evidence}
\end{figure*}

\subsection{Complementarity of Physical Base and\texorpdfstring{\\}{ }Neural Residual}
The representation diagnostic contains nine synthetic and six separate
measured-profile scenarios with enhanced noise/jitter. It compares removal of
the physical base or neural residual against a constant map and a 2D grid.
Table~\ref{tab:ablation} reports the mean, worst case, and per-scenario wins.
This diagnostic supports the representation choice; the following 12-case diagnostic benchmark
experiment tests the complete method.

\begin{table}[!t]
\centering
\caption{Comprehensive ablation summary over 15 representation scenarios (9 synthetic + 6 separate measured-profile cases). Values are PSNR (dB); underlining marks the best value in each summary column.}
\label{tab:ablation}
\scriptsize
\setlength{\tabcolsep}{3pt}
\begin{tabular}{lccc}
\toprule
\textbf{Variant} & \textbf{Avg} & \textbf{Worst} & \textbf{Wins} \\
\midrule
\textbf{Neural Grey} & \underline{31.31} & 16.11 & \underline{10/15} \\
w/o Physics Prior (Pure Neural) & 28.50 & 9.71 & 0/15 \\
w/o Neural Residual (Pure Physics) & 28.34 & \underline{18.36} & 5/15 \\
Remove All (Output 1.0) & 23.43 & 7.35 & 0/15 \\
Pure 2D Grid & 26.52 & 14.49 & 0/15 \\
\bottomrule
\end{tabular}
\end{table}

Combining the physical base and neural
residual gives the highest mean PSNR (31.31~dB) and wins 10/15 cases.
The physical-only model wins the other five cases and has a higher worst-case
score (18.36 versus 16.11~dB). Across the full study, adding the bounded neural
residual improves mean PSNR by 2.97~dB, while retaining the physical base
improves it by 2.81~dB over the neural-only variant. Together, these comparisons
support a constrained spatial correction around an optical model: the two
components provide complementary benefits to average recovery.

\begin{table*}[!t]
\centering
\caption{Radial selection, metadata, and capture-count ablation on the
12-case diagnostic benchmark. All entries are means except Worst and Wins; $\Delta$ is with
respect to the fixed four-capture full model. Selector/metadata rows use the
same joint fit; capture-count rows refit the first two, three, or four captures.
The two-/three-capture anchor is a pixelwise median; four captures
use a trimmed mean. Metrics use the fixed full-frame domain without
ground-truth scale fitting.}
\label{tab:selector_gauge_ablation}
\scriptsize
\setlength{\tabcolsep}{5.5pt}
\renewcommand{\arraystretch}{1.04}
\resizebox{0.88\textwidth}{!}{%
\begin{tabular}{lrrrrrr}
\toprule
\textbf{Variant} & \textbf{PSNR $\uparrow$} & \textbf{$\Delta$ (dB)} &
\textbf{SSIM $\uparrow$} & \textbf{$R^2\uparrow$} &
\textbf{Worst $\uparrow$} & \textbf{Wins} \\
\midrule
Joint fit only & 24.28 & -4.54 & 0.9829 & 0.730 & 17.87 & -- \\
$+D_2$ stabilization & 24.65 & -4.17 & 0.9839 & 0.740 & 17.55 & -- \\
$+$ radial selector (no $D_2$) & 27.12 & -1.69 & 0.9913 & 0.816 & 19.99 & -- \\
\textbf{Full model} & \textbf{28.82} & \textbf{0.00} & \textbf{0.9932} & \textbf{0.838} & \textbf{20.12} & -- \\
No focal metadata ($D_2$ only) & 24.65 & -4.17 & 0.9839 & 0.740 & 17.55 & -- \\
\midrule
2 captures ($0^\circ/90^\circ$) & 28.70 & -0.12 & 0.9929 & 0.783 & 19.63 & 4/12 \\
3 captures ($0^\circ/90^\circ/180^\circ$) & 28.54 & -0.27 & 0.9930 & 0.789 & 19.75 & 1/12 \\
4 captures ($0^\circ/90^\circ/180^\circ/270^\circ$) & 28.82 & 0.00 & 0.9932 & 0.838 & 20.12 & 7/12 \\
\bottomrule
\end{tabular}}

\end{table*}

\subsection{Role of Focal Metadata and Radial Selection}
Table~\ref{tab:selector_gauge_ablation} tests the contribution of radial
selection. The joint fit alone reaches 24.28~dB;
$D_2$ stabilization adds 0.37~dB. Applying the radial selector directly
to the joint estimate yields 27.12~dB, and preceding it with $D_2$
stabilization supplies another 1.69~dB to reach 28.82~dB. Thus
radial selection provides the largest gain, while $D_2$ provides
a complementary stabilization.

\subsection{Effect of Capture Count and Orientation Sampling}
Mean PSNR does not increase with every additional capture: two, three, and
four captures obtain 28.70, 28.54, and 28.82~dB. Four captures win 7/12 cases,
compared with 4/12 and 1/12 for two and three captures, and attain the highest
mean and best worst-case score (20.12 versus 19.63/19.75~dB).
The four captures sample evenly spaced camera orientations at
$0^\circ/90^\circ/180^\circ/270^\circ$.
Two captures remain best in four cases, so the result supports a default
protocol rather than a minimum capture requirement.

\subsection{Robustness to Capture Perturbations and\texorpdfstring{\\}{ }Hyperparameters}
We test capture errors by varying one acquisition factor at a time
on the 12-case diagnostic benchmark, using three scene seeds per case. At the strongest
tested perturbations, mean PSNR differs from the zero level for that factor
by less than 0.056~dB (Table~\ref{tab:tolerance}). The Supplementary Material gives
the capture-generation settings.

\begin{table}[t]
\centering
\caption{One-factor capture tolerance on the 12-case diagnostic benchmark.
PSNR averages 12 cases and three seeds at the strongest endpoint;
$\Delta$ compares with the zero level of the same sweep.}
\label{tab:tolerance}
\scriptsize
\setlength{\tabcolsep}{5pt}
\begin{tabular}{lrrr}
\toprule
\textbf{Factor} & \textbf{Swept $\sigma$} &
\textbf{PSNR} & \textbf{$\Delta$ vs. 0} \\
\midrule
Roll & $0$--$3^\circ$ & 28.618 & $-0.016$ \\
Translation & $0$--$8$ px & 28.587 & $-0.055$ \\
Scale & $0$--$3\%$ & 28.669 & $+0.041$ \\
Log exposure & $0$--$0.15$ & 28.628 & $-0.005$ \\
\bottomrule
\end{tabular}
\end{table}

For numerical sensitivity, we vary each loss weight
by a factor of two and test nearby network sizes and representation settings
while holding the other settings fixed. Changing the residual cap to
0.12/0.24, removing harmonics, or simplifying illumination gives mean-PSNR
spans of 0.039~dB before and 0.036~dB after radial selection. On the 12-case diagnostic benchmark under the
50-epoch protocol, network-size spans reach 0.078~dB, while the largest loss
weight span is 0.473~dB for the residual penalty
(Fig.~\ref{fig:design_evidence}c). The small capacity effect supports compact
fixed networks; the larger residual-weight effect shows that regularization
still matters.

Selector constants are tested separately on Enhanced-36, a Lensfun-based
benchmark with added noise and pose perturbations. Most spans are below 0.15~dB, but changing the wide-profile cosine exponent
produces a 0.612~dB span. This exponent directly controls the strength of the
wide-angle radial prior, making that choice more consequential than binning
or smoothing within the tested ranges. The sensitivity results therefore
support compact fixed fields while identifying radial-prior strength and
residual regularization as the more influential choices.

\section{Evaluation Across Optical Configurations}

We evaluate four complementary benchmarks: Synthetic-100 contains synthetic
shading and illumination; RayTraced-12 contains simulated optical
architectures; Real-21 contains real calibration takes; and Enhanced-36
combines measured Lensfun profiles with simulated capture perturbations.
We compare with Zheng et al.~\cite{zheng2009vignetting},
Cao et al.~\cite{zhang2022gaussian}, Goldman~\cite{goldman2010vignette},
Kim and Pollefeys~\cite{kim2008vignetting},
DeVigNet (4-Shot)~\cite{devignet2024},
Abdelhamed et al.~\cite{abdelhamed2022filters}, CNN-4Shot,
UNet-4Shot~\cite{chen2018cnn}, Analytic Radial (1F) and Analytic Radial (4F), and
VCLUT-inspired~\cite{vclut2023} baselines.
The 1F/4F suffixes denote single- and four-frame variants; DeVigNet (1-Shot)
is additionally reported on Synthetic-100. We report PSNR, SSIM \cite{wang2004image}, and MAE in one
fixed full-frame domain. Synthetic and ray-traced profiles lack physical
focal/sensor pairs and therefore test the joint fit; real-camera benchmarks
test the complete architecture. Single-frame methods use the canonical
synthetic/ray capture and the mean of four independent estimates on Real-21 and
Enhanced-36; multi-capture methods receive the same four images. The Supplementary Material defines each adaptation and gives the per-take protocol.
The learned comparisons cover fixed pretrained inference (DeVigNet (4-Shot)),
sparse per-instance fitting (CNN-4Shot and UNet-4Shot), lookup-table modeling (VCLUT-inspired),
and the Abdelhamed Gaussian-family adapter.

\begin{table}[!t]
\centering
\caption{Synthetic-100. Average PSNR/SSIM/MAE; $N$ is the number of evaluated cases. Underlining marks the best value in each metric column, including ties.}
\label{tab:synthetic}
\vspace{0.25em}
\scriptsize
\setlength{\tabcolsep}{4pt}
\begin{tabular}{lcccc}
\toprule
\textbf{Model} & \textbf{$N$} & \textbf{PSNR$\uparrow$} & \textbf{SSIM$\uparrow$} & \textbf{MAE$\downarrow$} \\
\midrule
\textbf{Neural Grey} & 100 & \underline{25.43} & \underline{1.00} & \underline{0.06} \\
Zheng et al. \cite{zheng2009vignetting} & 100 & 13.58 & 0.96 & 0.18 \\
Goldman \cite{goldman2010vignette} & 100 & 12.53 & 0.96 & 0.20 \\
Kim and Pollefeys \cite{kim2008vignetting} & 100 & 17.88 & 0.98 & 0.11 \\
Cao et al. \cite{zhang2022gaussian} & 100 & 17.31 & 0.92 & 0.14 \\
DeVigNet (4-Shot) \cite{devignet2024} & 100 & 23.72 & \underline{1.00} & \underline{0.06} \\
DeVigNet (1-Shot) \cite{devignet2024} & 100 & 16.08 & 0.91 & 0.14 \\
Abdelhamed et al.\ \cite{abdelhamed2022filters} & 100 & 16.50 & 0.98 & 0.12 \\
CNN-4Shot & 100 & 15.60 & 0.96 & 0.16 \\
UNet-4Shot \cite{chen2018cnn} & 100 & 13.48 & 0.95 & 0.21 \\
\mbox{Analytic Radial (4F)} & 100 & 15.88 & 0.93 & 0.16 \\
VCLUT-inspired \cite{vclut2023} & 100 & 16.79 & 0.98 & 0.12 \\
\mbox{Analytic Radial (1F)} & 100 & 17.03 & 0.97 & 0.13 \\
\bottomrule
\end{tabular}
\end{table}

\subsection{Synthetic Shading and Illumination}
Synthetic-100 spans ten degradation categories across 100 cases. Shading comes from
cosine-power, even-polynomial, non-monotonic multi-zone, decentered elliptical,
and nonparametric spline families. Illumination ranges from smooth gradients to
hard edges and high-frequency texture. Joint perturbations reach standard
deviations of $2.2^\circ$ roll, 14-pixel translation, 1.4\% scale, and 5\%
exposure, with photo-response non-uniformity (PRNU), read, and shot noise. Every baseline covers all 100 cases.

Neural Grey ranks first at 25.43~dB, 1.70~dB above DeVigNet (4-Shot). The
strongest classical result is Kim and Pollefeys at 17.88~dB. Category-level tables in the Supplementary Material show that physical regularization is most useful under the
edge-extreme and chaotic-asymmetry conditions.

\subsection{Validation on Ray-Traced Optical Architectures}

To verify physically plausible shading behavior, RayTraced-12 evaluates 12 classic mathematical lens architectures (e.g., Singlet, Cooke Triplet, Double Gauss, Cassegrain reflector) derived from ray-tracing simulations \cite{smith2004modern}. Table~\ref{tab:rayoptics} summarizes the architecture-level comparison.

\begin{table}[!t]
\centering
\caption{RayTraced-12: 12 optical architectures.
Neural Grey's 24.49~dB is the finite-value mean over 11 cases because one
zero-error case has infinite PSNR; SSIM and MAE retain all 12. Underlining marks
the best value in each metric column, including ties.}
\label{tab:rayoptics}
\scriptsize
\setlength{\tabcolsep}{4pt}
\begin{tabular}{lcccc}
\toprule
\textbf{Model} & \textbf{$N$} & \textbf{PSNR$\uparrow$} & \textbf{SSIM$\uparrow$} & \textbf{MAE$\downarrow$} \\
\midrule
\textbf{Neural Grey} & 12 & \underline{24.49} & \underline{0.90} & \underline{0.06} \\
Zheng et al. \cite{zheng2009vignetting} & 12 & 16.71 & 0.80 & 0.14 \\
Cao et al. \cite{zhang2022gaussian} & 12 & 18.99 & 0.89 & 0.11 \\
Goldman \cite{goldman2010vignette} & 12 & 10.32 & 0.75 & 0.35 \\
Kim and Pollefeys \cite{kim2008vignetting} & 12 & 21.95 & 0.89 & 0.09 \\
DeVigNet (4-Shot) \cite{devignet2024} & 12 & 14.18 & 0.78 & 0.20 \\
Abdelhamed et al.\ \cite{abdelhamed2022filters} & 12 & 17.10 & 0.82 & 0.18 \\
CNN-4Shot & 12 & 17.72 & 0.87 & 0.14 \\
UNet-4Shot \cite{chen2018cnn} & 12 & 17.28 & 0.82 & 0.15 \\
\mbox{Analytic Radial (4F)} & 12 & 20.51 & 0.80 & 0.14 \\
VCLUT-inspired \cite{vclut2023} & 12 & 10.07 & 0.74 & 0.32 \\
\mbox{Analytic Radial (1F)} & 12 & 15.82 & 0.84 & 0.20 \\
\bottomrule
\end{tabular}
\end{table}

Neural Grey's finite-value ray-traced mean is 24.49~dB, ahead of
Kim and Pollefeys at 21.95~dB by +2.54~dB and Analytic Radial (4F) at
20.51~dB. On the simulated Cassegrain stress profile, whose annular response
violates the assumption that the optical center is brightest, Neural Grey
reaches 17.00~dB while DeVigNet (4-Shot) reaches 12.99~dB. Together with the
commercial-lens benchmark below, these simulations cover diverse optical structure.

\begin{table*}[!t]
\centering
\caption{Real-21: seven commercial optical configurations, three takes each. Values
are absolute PSNR (dB), with no per-prediction scalar alignment; underlining
marks the best result per column. Parentheses identify the tested zoom setting.
All configurations were captured at maximum aperture: from left to right,
$f/1.4$, $f/3.5$, $f/5.6$, $f/1.2$, $f/2.8$, $f/2.5$, and $f/2.8$.
The Nikon 28--80~mm is the AF-D version.}
\label{tab:real_world}
\scriptsize
\setlength{\tabcolsep}{2.5pt}
\renewcommand{\arraystretch}{1.12}
\begin{tabular}{@{}lcccccccc@{}}
\toprule
\textbf{Model} & \shortstack{\textbf{Sigma}\\\textbf{20~mm}} &
\shortstack{\textbf{Nikon 28--80~mm}\\\textbf{(28~mm)}} &
\shortstack{\textbf{Nikon 28--80~mm}\\\textbf{(80~mm)}} &
\shortstack{\textbf{Canon FL}\\\textbf{55~mm}} &
\shortstack{\textbf{Nikon Micro}\\\textbf{60~mm}} &
\shortstack{\textbf{Nikon}\\\textbf{105~mm}} &
\shortstack{\textbf{Nikon 80--200~mm}\\\textbf{(200~mm)}} & \textbf{Avg.} \\
\midrule
\textbf{Neural Grey} & \underline{26.17} & \underline{23.85} &
\underline{22.43} & 20.87 & 12.98 & \underline{35.41} &
\underline{35.95} & \underline{25.38} \\
Kim and Pollefeys \cite{kim2008vignetting} & 12.15 & 14.16 & 15.43 &
13.46 & 13.23 & 21.70 & 15.78 & 15.13 \\
Cao et al. \cite{zhang2022gaussian} & 12.04 & 12.42 & 20.35 &
\underline{21.32} & 12.56 & 25.82 & 33.84 & 19.76 \\
Zheng et al. \cite{zheng2009vignetting} & 12.28 & 12.57 & 19.71 &
21.07 & 12.44 & 26.95 & 32.99 & 19.72 \\
\mbox{Analytic Radial (1F)} & 11.47 & 13.36 & 19.57 & 19.85 &
12.48 & 25.69 & 31.51 & 19.13 \\
Goldman \cite{goldman2010vignette} & 3.52 & 7.18 & 10.34 & 7.43 &
7.63 & 15.82 & 9.03 & 8.71 \\
DeVigNet (4-Shot) \cite{devignet2024} & 9.65 & 10.38 & 21.09 & 18.48 &
12.70 & 12.15 & 18.81 & 14.75 \\
Abdelhamed et al.\ \cite{abdelhamed2022filters} & 12.85 & 12.01 & 20.63 &
19.93 & 12.15 & 20.98 & 31.92 & 18.64 \\
VCLUT-inspired \cite{vclut2023} & 7.33 & 12.93 & 14.29 & 16.15 &
11.13 & 21.80 & 15.15 & 14.11 \\
CNN-4Shot & 11.84 & 16.39 & 21.48 & 13.75 & 13.25 & 22.91 & 27.25 &
18.13 \\
UNet-4Shot \cite{chen2018cnn} & 13.00 & 16.43 & 17.95 & 14.04 &
\underline{13.28} & 14.47 & 21.81 & 15.85 \\
\bottomrule
\end{tabular}
\end{table*}

\begin{figure*}[!t]
\centering
\includegraphics[width=0.92\textwidth]{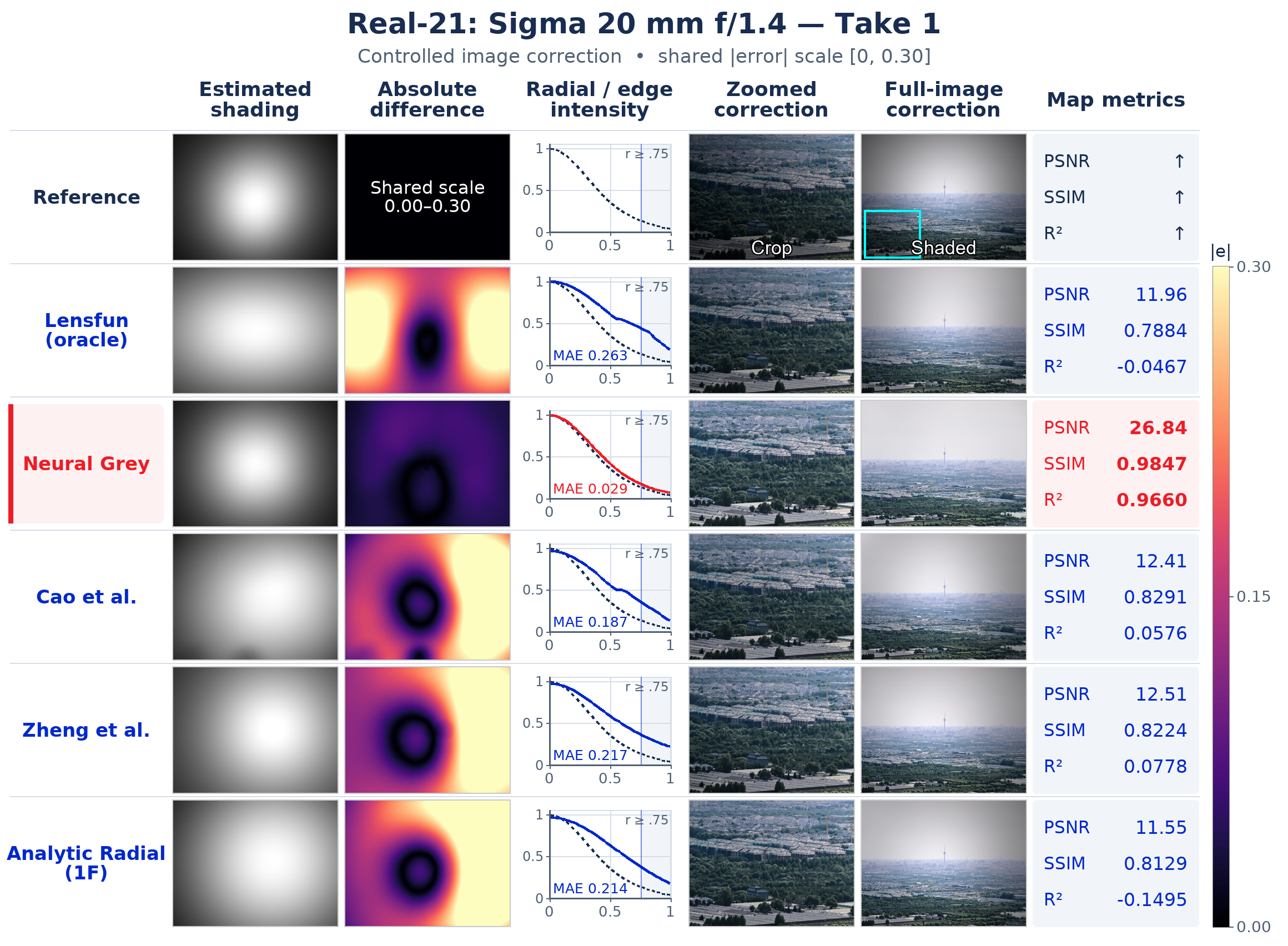}\par
\caption{Real-21 Sigma 20~mm f/1.4, Take 1. Difference maps share a fixed
$[0,0.30]$ linear-intensity scale. The textured-image panels are controlled
simulations: measured shading is applied in linear RGB and removed using each
estimate. Map metrics use the displayed 1000-pixel estimates and the independent
flat-field reference; the quantitative tables use a 1024-pixel long edge.
Single-frame qualitative baselines use the first capture. Lensfun uses interpolation selected using ground truth (Sec.~\ref{sec:lensfun}).}
\label{fig:real_world_sigma20_take1}

\end{figure*}

\begin{table*}[!t]
\centering
\caption{Enhanced-36. Each lens column is the mean absolute PSNR (dB) over six fixed wall/illumination configurations; all 36 cases use the Table~\ref{tab:real_world} protocol. Underlining marks the best result per column.}
\label{tab:rwlf36}
\footnotesize
\setlength{\tabcolsep}{2pt}
\renewcommand{\arraystretch}{1.04}
\resizebox{\textwidth}{!}{%
\begin{tabular}{lccccccc}
\toprule
\textbf{Model} & \shortstack{\textbf{iPhone XS}\\\textbf{4.25~mm}} &
\shortstack{\textbf{Sigma 19~mm}\\\textbf{$f/2.8$ DN}} &
\shortstack{\textbf{Pentax-DA}\\\textbf{18--135~mm}\\\textbf{(100~mm)}} &
\shortstack{\textbf{Sigma 60--600~mm}\\\textbf{$+1.4\times$ converter}\\\textbf{(840~mm)}} &
\shortstack{\textbf{Laowa 85~mm}\\\textbf{$2\times$ Ultra Macro}} &
\shortstack{\textbf{Olympus 75~mm}\\\textbf{$f/1.8$}} & \textbf{Avg.} \\
\midrule
\textbf{Neural Grey} & \underline{25.02} & \underline{21.11} & \underline{17.81} & 18.97 & \underline{20.62} & \underline{20.06} & \underline{20.60} \\
Kim and Pollefeys \cite{kim2008vignetting} & 20.09 & 16.82 & 17.57 & 17.59 & 17.02 & 17.94 & 17.84 \\
Cao et al. \cite{zhang2022gaussian} & 18.50 & 14.60 & 10.12 & 13.32 & 13.00 & 11.77 & 13.55 \\
Zheng et al. \cite{zheng2009vignetting} & 19.42 & 15.19 & 10.48 & 14.64 & 13.63 & 12.36 & 14.28 \\
Analytic Radial (4F) & 15.27 & 9.48 & 12.36 & 12.38 & 12.70 & 6.57 & 11.46 \\
Goldman \cite{goldman2010vignette} & 6.51 & 11.52 & 15.80 & 15.98 & 13.16 & 15.29 & 13.04 \\
DeVigNet (4-Shot) \cite{devignet2024} & 15.70 & 17.94 & 12.55 & 16.19 & 17.67 & 12.93 & 15.50 \\
Abdelhamed et al. \cite{abdelhamed2022filters} & 19.93 & 17.66 & 17.18 & 17.94 & 15.29 & 18.51 & 17.75 \\
VCLUT-inspired \cite{vclut2023} & 12.64 & 17.47 & 16.36 & \underline{19.11} & 17.13 & 17.34 & 16.68 \\
CNN-4Shot & 19.86 & 14.14 & 9.85 & 12.36 & 12.17 & 11.44 & 13.30 \\
UNet-4Shot \cite{chen2018cnn} & 20.76 & 12.32 & 9.10 & 11.19 & 11.11 & 9.97 & 12.41 \\
\bottomrule
\end{tabular}}
\end{table*}

\subsection{Real-World Performance on Commercial Lenses}

Real-21 contains 21 hand-held takes from seven Sigma, Canon, and Nikon
configurations spanning prime, macro, wide-angle, and telephoto designs. Each
take comprises four calibration captures. Ground
truth maps come from separate controlled flat-field captures and are never
available to the estimator. Each reference RAW is linearized per Bayer site
using its recorded black and white levels, combined as linear luminance,
impulse-suppressed, Gaussian-smoothed at 1.2\% of the long edge, and normalized
by a robust central-peak estimate. The four casual calibration captures and the
reference use the same camera--lens--aperture--focus configuration and share no
frames. Every value in Table~\ref{tab:real_world} is the
three-take mean absolute PSNR, without post-hoc scalar alignment. With a
1024-pixel long edge, Neural Grey reaches
25.38~dB, exceeding the strongest competing mean (Cao et al., 19.76~dB) by
5.62~dB. This margin is not driven by the high-PSNR telephoto cases: after
removing all six 105 and 200~mm takes, Neural Grey still averages 21.26~dB on
the remaining 15 takes versus 15.74~dB for the strongest remaining competitor
(Cao et al.), a $+5.52$~dB margin. It ranks first on five of seven configurations, including the Sigma
20~mm and Nikon 28~mm wide-angle settings as well as the 80~mm, 105~mm,
and 200~mm settings. On Canon FL 55~mm, Neural Grey reaches 20.87~dB, 0.45~dB below
Cao et al. (21.32~dB); on Micro 60~mm, it reaches 12.98~dB, 0.30~dB below
UNet-4Shot (13.28~dB). Micro 60~mm yields low absolute PSNR across methods.
A diagnostic scalar correction removes little of Neural Grey's error,
indicating that the remaining error is mainly spatial (see the Supplementary Material).

Neural Grey is outright best on 14/21 takes and wins 19/21 paired
comparisons against Cao et al., the strongest competitor by full-benchmark mean, and
18/21 against Zheng et al. Repeated takes
test capture variability, while Enhanced-36 broadens the optical and capture
conditions through measured Lensfun profiles with added noise and jitter.
The per-take results are retained in the Supplementary Material.

Fig.~\ref{fig:real_world_sigma20_take1} shows the strongest Neural Grey Sigma
20~mm take (26.88~dB in the per-take results; 26.84~dB at the figure's
1000-pixel resolution) together with Lensfun and the
three strongest non-Neural-Grey methods by full Real-21 average PSNR.
A separate controlled textured-image experiment applies each measured shading
map to one common clean photograph, then corrects it using the estimated map.
Across the 21 takes, the complete estimator reaches 31.40~dB, compared with
29.00~dB for the joint fit and 19.58~dB without correction (see the Supplementary Material).
The main evaluation target is recovery of a reusable shading map against an
independent flat-field reference; this controlled image experiment illustrates
how map accuracy translates into brightness correction.

\subsection{Comparison with Standard LUT Software}
\label{sec:lensfun}
Lensfun represents the standard metadata-driven LUT correction available in
photographic software: a lens identity, focal length, aperture, and focus
distance select an interpolated lens profile from the database. For each
covered lens, flat-field ground truth selects the best interpolated Lensfun
profile. Across exact Sigma 20~mm and
Micro 60 AF-D records, an optically equivalent AF/AF-D 80--200~mm record, and a
Nikon 105~mm generation candidate, even this oracle-selected LUT remains below Neural
Grey's in-situ estimate. Canon FL 55~mm has no database profile. The complete
values are in the Supplementary Material. This demonstrates that our in-situ,
configuration-specific estimation outperforms standard database LUTs, even
with oracle profile selection.

\subsection{Enhanced-36}
Enhanced-36 uses the six Lensfun lens profiles in Table~\ref{tab:rwlf36},
each under six simulated capture conditions.
Every case combines a measured Lensfun shading profile with an irregular
Lambertian wall, separate illumination, 0.5--3\% half-diagonal decentering,
small roll/pose and exposure variation, plus shot/read/row noise.
Table~\ref{tab:rwlf36} evaluates all
Table~\ref{tab:real_world} methods together with Abdelhamed et al. using the same
comparison procedure.

Neural Grey leads five of the six lens columns and the 36-case average at
20.60~dB. Kim and Pollefeys' method ranks second at 17.84~dB, followed by
Abdelhamed et al. at 17.75~dB. On the remaining configuration, a Sigma
60--600~mm with a 1.4$\times$ teleconverter at 840~mm, Neural Grey reaches
18.97~dB, within 0.14~dB of VCLUT-inspired (19.11~dB). Across the six optical
profiles and their capture perturbations, the same fixed configuration
achieves the highest average and leads five lens groups, supporting the
method's applicability across the tested optical conditions.

\newpage
\section{Conclusion}
Neural Grey separates sensor-fixed shading from scene-fixed illumination by
combining four-roll radiometric factorization, a bounded physics--neural field,
and focal metadata to select the otherwise ambiguous radial profile. It ranks first on
the synthetic and ray-traced benchmarks and reaches 25.38~dB over 21 commercial-lens
takes, 5.62~dB above the strongest competing real-world average. The same
24.8K-parameter formulation covers 20--200~mm configurations and produces a
reusable, unit-specific shading map.

\begingroup
\small
\bibliographystyle{ieee_fullname}

\begin{thebibliography}{99}
\footnotesize
\linespread{0.98}\selectfont
\setlength{\itemsep}{0pt}
\setlength{\parsep}{0pt}
\setlength{\parskip}{0pt}

\bibitem{debevec1997hdr}
P.~E.~Debevec and J.~Malik, ``Recovering HDR radiance maps,'' in {\em SIGGRAPH}, 1997.

\bibitem{smith2015cidre}
K.~Smith, Y.~Li, F.~Piccinini, G.~Csucs, C.~Balazs, A.~Bevilacqua, and P.~Horvath,
\newblock ``CIDRE: An illumination-correction method for optical microscopy,''
\newblock {\em Nature Methods}, vol.~12, pp.~404--406, 2015.

\bibitem{hecht2016optics}
E.~Hecht,
\newblock {\em Optics}, 5th ed.
\newblock Boston, MA, USA: Pearson, 2016.

\bibitem{cooke2021itech}
Cooke Optics Ltd.,
\newblock ``/i Technology Reference Manual: Intelligent Lens Metadata Protocol,''
\newblock Tech. Rep., 2021.

\bibitem{zheng2009vignetting}
Y.~Zheng, S.~Lin, and S.~B.~Kang,
\newblock ``Single-image vignetting correction,''
\newblock {\em IEEE TPAMI}, 2009.

\bibitem{yu2004vignetting}
W.~Yu, ``Practical anti-vignetting methods,'' {\em IEEE TCE}, 2004.

\bibitem{kask2016flatfield}
P.~Kask, K.~Palo, C.~Hinnah, and T.~Pommerencke,
\newblock ``Flat field correction for high-throughput imaging of fluorescent samples,''
\newblock {\em J. Microscopy}, vol.~263, no.~3, pp.~328--340, 2016.

\bibitem{kang2000vignetting}
S.~B.~Kang and R.~Weiss,
\newblock ``Can we calibrate a camera using an image of a flat, textureless Lambertian surface?,''
\newblock in {\em ECCV}, 2000.

\bibitem{kim2008vignetting}
S.~J.~Kim and M.~Pollefeys,
\newblock ``Robust radiometric calibration and vignetting correction,''
\newblock {\em IEEE TPAMI}, vol.~30, no.~4, pp.~562--576, 2008.

\bibitem{goldman2010vignette}
D.~B.~Goldman,
\newblock ``Vignette and exposure calibration and compensation,''
\newblock {\em IEEE TPAMI}, 2010.

\bibitem{likar2000entropy}
B.~Likar, J.~B.~A.~Maintz, M.~A.~Viergever, and F.~Pernu\v{s},
\newblock ``Retrospective shading correction based on entropy minimization,''
\newblock {\em J. Microscopy}, vol.~197, no.~3, pp.~285--295, 2000.

\bibitem{peng2017basic}
T.~Peng, K.~Thorn, T.~Schroeder, L.~Wang, F.~J.~Theis, C.~Marr, and N.~Navab,
\newblock ``A BaSiC tool for background and shading correction of optical microscopy images,''
\newblock {\em Nature Communications}, vol.~8, Art.~no.~14836, 2017.

\bibitem{mitsunaga1999radiometric}
T.~Mitsunaga and S.~K.~Nayar,
\newblock ``Radiometric self calibration,''
\newblock in {\em Proc. IEEE CVPR}, vol.~1, pp.~374--380, 1999.

\bibitem{grossberg2003knowable}
M.~D.~Grossberg and S.~K.~Nayar,
\newblock ``Determining the camera response from images: What is knowable?,''
\newblock {\em IEEE TPAMI}, vol.~25, no.~11, pp.~1455--1467, 2003.

\bibitem{mo2017radiometric}
Z.~Mo, B.~Shi, S.-K.~Yeung, and Y.~Matsushita,
\newblock ``Radiometric calibration for Internet photo collections,''
\newblock in {\em Proc. IEEE CVPR}, pp.~4738--4746, 2017.

\bibitem{li2017faces}
C.~Li, S.~Lin, K.~Zhou, and K.~Ikeuchi,
\newblock ``Radiometric calibration from faces in images,''
\newblock in {\em Proc. IEEE CVPR}, pp.~3117--3126, 2017.

\bibitem{ronneberger2015unet}
O.~Ronneberger, P.~Fischer, and T.~Brox,
\newblock ``U-Net: Convolutional networks for biomedical image segmentation,''
\newblock in {\em MICCAI}, 2015.

\bibitem{chen2018cnn}
C.~Chen, Q.~Chen, J.~Xu, and V.~Koltun,
\newblock ``Learning to see in the dark,''
\newblock in {\em CVPR}, 2018.

\bibitem{vclut2023}
Y.~Li et al.,
\newblock ``Prior Visual-Guided Self-Supervised Learning Enables Color Vignetting Correction,''
\newblock {\em IEEE JBHI}, 2024.

\bibitem{devignet2024}
S.~Luo et al.,
\newblock ``DeVigNet: High-Resolution Vignetting Removal via a Dual Aggregated Fusion Transformer,''
\newblock in {\em AAAI}, 2024.

\bibitem{abdelhamed2022filters}
A.~Abdelhamed, J.~Yim, A.~Punnappurath, M.~S.~Brown, J.~Choe, and K.~Kim,
\newblock ``Extracting vignetting and grain filter effects from photos,''
\newblock in {\em Proc. IEEE/CVF WACV}, 2022.

\bibitem{land1971retinex}
E.~H.~Land and J.~J.~McCann,
\newblock ``Lightness and Retinex theory,''
\newblock {\em J. Opt. Soc. Am.}, vol.~61, no.~1, pp.~1--11, 1971.

\bibitem{jobson1997multiscale}
D.~J.~Jobson, Z.~Rahman, and G.~A.~Woodell,
\newblock ``A multiscale Retinex for bridging the gap between color images and the human observation of scenes,''
\newblock {\em IEEE TIP}, vol.~6, no.~7, pp.~965--976, 1997.

\bibitem{wei2018retinex}
C.~Wei, W.~Wang, W.~Yang, and J.~Liu,
\newblock ``Deep Retinex decomposition for low-light enhancement,''
\newblock in {\em Proc. BMVC}, Art.~no.~155, 2018.

\bibitem{zhang2019kind}
Y.~Zhang, J.~Zhang, and X.~Guo,
\newblock ``Kindling the darkness: A practical low-light image enhancer,''
\newblock in {\em Proc. ACM Multimedia}, pp.~1632--1640, 2019.

\bibitem{guo2020zerodce}
C.~Guo, C.~Li, J.~Guo, C.~C.~Loy, J.~Hou, S.~Kwong, and R.~Cong,
\newblock ``Zero-reference deep curve estimation for low-light image enhancement,''
\newblock in {\em Proc. IEEE/CVF CVPR}, pp.~1780--1789, 2020.

\bibitem{liu2021ruas}
R.~Liu, L.~Ma, J.~Zhang, X.~Fan, and Z.~Luo,
\newblock ``Retinex-inspired unrolling with cooperative prior architecture search for low-light image enhancement,''
\newblock in {\em Proc. IEEE/CVF CVPR}, pp.~10561--10570, 2021.

\bibitem{ma2022sci}
L.~Ma, T.~Ma, R.~Liu, X.~Fan, and Z.~Luo,
\newblock ``Toward fast, flexible, and robust low-light image enhancement,''
\newblock in {\em Proc. IEEE/CVF CVPR}, pp.~5637--5646, 2022.

\bibitem{xu2022snr}
X.~Xu, R.~Wang, C.-W.~Fu, and J.~Jia,
\newblock ``SNR-aware low-light image enhancement,''
\newblock in {\em Proc. IEEE/CVF CVPR}, pp.~17714--17724, 2022.

\bibitem{raissi2019pinn}
M.~Raissi, P.~Perdikaris, and G.~E.~Karniadakis,
\newblock ``Physics-informed neural networks: A deep learning framework,''
\newblock {\em J. Comput. Phys.}, 2019.

\bibitem{karniadakis2021piml}
G.~E.~Karniadakis, I.~G.~Kevrekidis, L.~Lu, P.~Perdikaris, S.~Wang, and L.~Yang,
\newblock ``Physics-informed machine learning,''
\newblock {\em Nature Reviews Physics}, vol.~3, no.~6, pp.~422--440, 2021.

\bibitem{mildenhall2020nerf}
B.~Mildenhall, P.~P.~Srinivasan, M.~Tancik, J.~T.~Barron, R.~Ramamoorthi, and R.~Ng,
\newblock ``NeRF: Representing scenes as neural radiance fields for view synthesis,''
\newblock in {\em ECCV}, 2020.

\bibitem{sitzmann2020siren}
V.~Sitzmann, J.~N.~P.~Martel, A.~W.~Bergman, D.~B.~Lindell, and G.~Wetzstein,
\newblock ``Implicit neural representations with periodic activation functions,''
\newblock in {\em NeurIPS}, 2020.

\bibitem{park2019deepsdf}
J.~J.~Park, P.~Florence, J.~Straub, R.~Newcombe, and S.~Lovegrove,
\newblock ``DeepSDF: Learning continuous signed distance functions for shape representation,''
\newblock in {\em Proc. IEEE/CVF CVPR}, pp.~165--174, 2019.

\bibitem{mescheder2019occupancy}
L.~Mescheder, M.~Oechsle, M.~Niemeyer, S.~Nowozin, and A.~Geiger,
\newblock ``Occupancy networks: Learning 3D reconstruction in function space,''
\newblock in {\em Proc. IEEE/CVF CVPR}, pp.~4460--4470, 2019.

\bibitem{chen2021liif}
Y.~Chen, S.~Liu, and X.~Wang,
\newblock ``Learning continuous image representation with local implicit image function,''
\newblock in {\em Proc. IEEE/CVF CVPR}, pp.~8628--8638, 2021.

\bibitem{yang2023nerco}
S.~Yang, M.~Ding, Y.~Wu, Z.~Li, and J.~Zhang,
\newblock ``Implicit neural representation for cooperative low-light image enhancement,''
\newblock in {\em Proc. IEEE/CVF ICCV}, pp.~12918--12927, 2023.

\bibitem{ulyanov2018dip}
D.~Ulyanov, A.~Vedaldi, and V.~Lempitsky,
\newblock ``Deep image prior,''
\newblock in {\em Proc. IEEE CVPR}, pp.~9446--9454, 2018.

\bibitem{gandelsman2019doubledip}
Y.~Gandelsman, A.~Shocher, and M.~Irani,
\newblock ``Double-DIP: Unsupervised image decomposition via coupled deep-image-priors,''
\newblock in {\em Proc. IEEE/CVF CVPR}, pp.~11026--11035, 2019.

\bibitem{grossberg2004response}
M.~D.~Grossberg and S.~K.~Nayar,
\newblock ``Modeling the space of camera response functions,''
\newblock {\em IEEE TPAMI}, vol.~26, no.~10, pp.~1272--1282, 2004.

\bibitem{nam2017pipeline}
S.~Nam and S.~J.~Kim,
\newblock ``Modelling the scene dependent imaging in cameras with a deep neural network,''
\newblock in {\em Proc. IEEE ICCV}, pp.~1717--1725, 2017.

\bibitem{rahaman2019spectral}
N.~Rahaman, A.~Baratin, D.~Arpit, F.~Draxler, M.~Lin, F.~A.~Hamprecht, Y.~Bengio, and A.~Courville,
\newblock ``On the spectral bias of neural networks,''
\newblock in {\em ICML}, 2019.

\bibitem{tancik2020fourier}
M.~Tancik, P.~P.~Srinivasan, B.~Mildenhall, S.~Fridovich-Keil, N.~Raghavan, U.~Singhal, R.~Ramamoorthi, J.~T.~Barron, and R.~Ng,
\newblock ``Fourier features let networks learn high frequency functions in low dimensional domains,''
\newblock in {\em NeurIPS}, 2020.

\bibitem{lindell2022bacon}
D.~B.~Lindell, D.~Van~Veen, J.~J.~Park, and G.~Wetzstein,
\newblock ``BACON: Band-limited coordinate networks for multiscale scene representation,''
\newblock in {\em Proc. IEEE/CVF CVPR}, pp.~16252--16262, 2022.

\bibitem{huber1964robust}
P.~J.~Huber, ``Robust estimation,'' {\em Annals of Mathematical Statistics}, 1964.

\bibitem{kingma2015adam}
D.~P.~Kingma and J.~Ba,
\newblock ``Adam: A method for stochastic optimization,''
\newblock in {\em ICLR}, 2015.

\bibitem{loshchilov2017sgdr}
I.~Loshchilov and F.~Hutter,
\newblock ``SGDR: Stochastic gradient descent with warm restarts,''
\newblock in {\em ICLR}, 2017.

\bibitem{wang2004image}
Z.~Wang et al.,
\newblock ``Image quality assessment: from error visibility to structural similarity,''
\newblock {\em IEEE TIP}, 2004.

\bibitem{zhang2022gaussian}
H.~Cao, X.~Gu, M.~Zhang, H.~Zhang, and X.~Chen,
\newblock ``Vignetting Correction Based on a Two-Dimensional Gaussian Filter With Harmony for Area Array Sensors,''
\newblock {\em IEEE Transactions on Computational Imaging}, vol. 8, pp. 576-584, 2022.

\bibitem{smith2004modern}
W.~Smith,
\newblock {\em Modern Lens Design}.
\newblock McGraw-Hill Education, 2004.

\end{thebibliography}

\IEEEtriggeratref{34}

\endgroup

\end{document}